\documentclass[runningheads]{llncs}
\usepackage{graphicx}
\usepackage{subfig, booktabs, todonotes}
\newcommand\blfootnote[1]{%
  \begingroup
  \renewcommand\thefootnote{}\footnote{#1}%
  \addtocounter{footnote}{-1}%
  \endgroup
}
\begin{document}
\title{Global and Local Interpretability \\ for Cardiac MRI Classification}

\author{James R. Clough 
\and
Ilkay Oksuz
\and
Esther Puyol-Ant\'on
\and
Bram Ruijsink
\and \\
Andrew P. King
\and 
Julia A. Schnabel
}
\authorrunning{J. R. Clough et al.}
\institute{School of Biomedical Engineering \& Imaging Sciences, King\rq{}s College London, UK
\email{james.clough@kcl.ac.uk}\\
}
\maketitle              
\begin{abstract}
Deep learning methods for classifying medical images have demonstrated impressive accuracy in a wide range of tasks but often these models are hard to interpret, limiting their applicability in clinical practice. 
In this work we introduce a convolutional neural network model for identifying disease in temporal sequences of cardiac MR segmentations which is interpretable in terms of clinically familiar measurements. 
The model is based around a variational autoencoder, reducing the input into a low-dimensional latent space in which classification occurs.
We then use the recently developed `concept activation vector' technique to associate concepts which are diagnostically meaningful (eg. clinical biomarkers such as `low left-ventricular ejection fraction') to certain vectors in the latent space.
These concepts are then qualitatively inspected by observing the change in the image domain resulting from interpolations in the latent space in the direction of these vectors.
As a result, when the model classifies images it is also capable of providing naturally interpretable concepts relevant to that classification and demonstrating the meaning of those concepts in the image domain.
Our approach is demonstrated on the UK Biobank cardiac MRI dataset where we detect the presence of coronary artery disease.
\keywords{Interpretable ML  \and Cardiac MRI \and Coronary artery disease.}
\end{abstract}
\section{Introduction}
\label{sec:introduction}
Heart disease is the leading cause of death globally.
\blfootnote{This work was supported by an EPSRC programme Grant (EP/P001009/1) and the Wellcome EPSRC Centre for Medical Engineering at the School of Biomedical Engineering and Imaging Sciences, King’s College London (WT 203148/Z/16/Z). This research has been conducted using the UK Biobank Resource under Application Numbers 40119 and 17806. The GPU used in this research was generously donated by the NVIDIA Corporation.}
Cardiac magnetic resonance (CMR) is the gold-standard imaging tool for assessment and diagnosis of many serious forms of heart disease \cite{Petersen2016}.
As the performance of machine learning (ML) tools for image classification has improved in recent years \cite{Bernard2018}, the interest in the application of ML to the analysis of CMR images and volumes has grown.
Such systems have the potential to provide significant benefits to patients such as improved diagnostic quality and decreased time and cost of image analysis.
However, ML methods successfully demonstrated in a research setting can face barriers to clinical application due to concerns about reliability and a lack of interpretability.
In particular, deep convolutional neural networks (CNN) have proven powerful tools for image analysis but their ability to yield adequate explanations of their decisions to clinicians is still lacking.
Interpretable ML models are important in healthcare for the trust of patients and clinicians, to guard against model unreliability in the face of distributional shift \cite{Patel2015} (e.g. due to a change in scanner design, imaging protocol and pre-processing, or patient demographics) and for legal reasons such as a patient's `right to explanation' \cite{Goodman2016}.

In this work we develop a classification framework using variational autoencoders (VAE) \cite{Kingma2013} which allows for both local and global interpretability of a classification decision.
By \emph{local interpretability} we mean the ability to ask 
`which features of this particular image led to it being classified in this particular way?'.
By \emph{global interpretability} we mean the ability to ask
`which common features were generally associated with images assigned to this particular class?'.
Our method first encodes 2D image segmentations into a low-dimensional latent space with a VAE and then classifies using the latent vectors.
Using concept activation vectors \cite{Kim2017a} in the space of activations in the intermediate layers of the classification network provides global interpretability to the model.
The VAE contains a decoder which is trained to reconstruct images from the latent vectors and so local interpretability is provided by interpolating in the latent space and visualising the changes in the corresponding decoded images.
This approach is demonstrated using cardiac segmentations, obtained from CMR studies in the UK Biobank, and classifying for the presence of coronary artery disease.
Our primary contribution in this work is the integration of local and global interpretability methods in the context of a realistic clinical application. 
Additionally, our proposed classification method utilises temporal information over the full cardiac cycle. 
This is important as dynamic features, such as regional and global myocardial wall motion, are sensitive markers of disease that are missed when only taking into account images at end-diastolic and/or end-systolic positions.
\section{Related Work}
\label{sec:related}
The importance of providing interpretability to image classification models is reflected in the growing body of literature around the subject.
Some classification models
such as simple decision trees or linear models
are considered to be inherently interpretable in the sense that a human observer can understand each step in the process by which a model makes a decision.
Unfortunately many ML models which have the most impressive classification performance 
and so are most desirable to use in clinical practice do not have this property.

When a model is too complex for its entire decision process to be understood, interpretability is still possible by supplementing the output decision with 
information which can help to explain it.
Saliency maps 
\cite{Simonyan2013}
are a commonly used approach for interpreting image classification in which the gradient of the loss with respect to the input image is visualised.
Although saliency maps can be useful for highlighting relevant regions of images, the level of interpretability that they can provide is often of limited use.
Firstly, as explained in \cite{Adebayo2018a}, \emph{`some widely deployed saliency methods are independent of both the data the model was trained on, and the model parameters'}, which is clearly undesirable.
Secondly, as noted in \cite{Rudin2018}, saliency methods only explain `where the network is looking'.
If an image of a dog is misclassified as a cat, and a saliency map highlights the region of the image containing the dog, we still do not know much about why this image was misclassified.
Thirdly, the explanation is only relevant for the particular image in question and so an observer must manually assess many images and their saliency maps to draw more general conclusions \cite{Kim2017a}.

Another family of approaches attempts to understand the representations learned by intermediate layers in a deep CNN by visualising the images which strongly activate each neuron 
\cite{Olah2018}.
While these methods are helpful for achieving a better understanding of how CNNs work, the images produced  
do not typically appear realistic and so are often hard to interpret themselves, appearing to capture textures more strongly than wider-scale structure.

Autoencoders are neural networks trained to find efficient representations of a dataset.
They do this with an encoder network, which maps images to low-dimensional latent vectors, and a decoder network which approximates the original image from the latent vector.
The representations learned by such models can be used to de-noise images 
or impose prior knowledge about allowed structures \cite{Oktay2018}.
In \cite{Biffi2018} a classification task (detecting hypertrophic cardiomyopathy from CMR volume segmentations) was performed in the latent space of a variational autoencoder.
This allowed the classifier to be understood because one can take the latent vector corresponding to a patient's CMR data, and interpolate it in the direction of the gradient of the classifier's output, and observe the changes to the decoded image.
Our method extends this autoencoder approach to use the whole cardiac cycle rather than just two frames, providing local interpretability, and integrates it with concept activation vectors \cite{Kim2017a}, a method for global interpretability.

\section{Methods}
\label{sec:methods}
\begin{figure}[bt]%
\noindent\makebox[\textwidth]{
    \centering
    \subfloat{{\includegraphics[width=\textwidth]{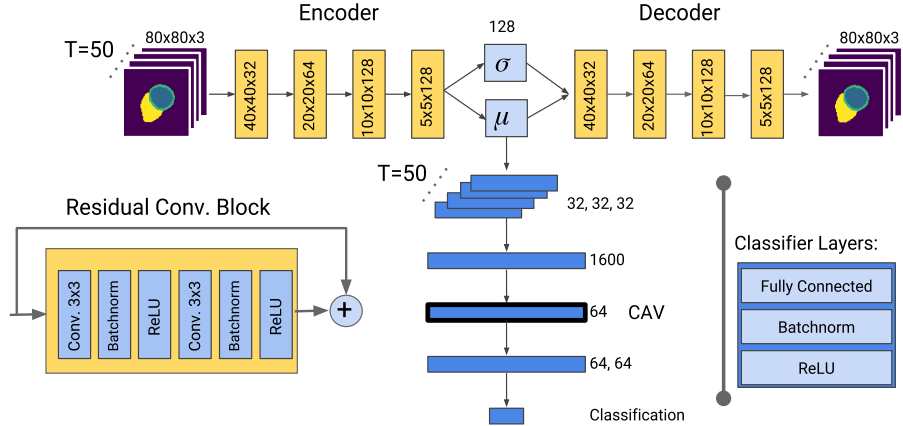}}}}%
    \caption{Diagram showing the architecture of the joint VAE/classification model.
    The VAE consists of a series of residual convolutional blocks, with the image resolution and number of feature maps denoted in each block.
    The classification network consists of a series of fully connected layers (number of hidden units in each denoted to the side) which first processes the latent vectors individually, then concatenates them and processes them together.}%
    \label{fig:model}%
\end{figure}

\textbf{VAE/Classification network:}
Our classification model is described by the diagram in Figure \ref{fig:model}.
The model consists firstly of an encoder which finds a latent representation (of dimensionality 128) for each input.
In our case, these inputs are $80 \times 80$ segmentations of 3 central slices in the stack of short-axis CMR images of the heart, where each slice is treated as a channel in the image.
A decoder network is trained to reconstruct the original data from the latent representations.
The data for each subject consists of $T=50$ segmentations per slice, representing one full cardiac cycle.
These segmentations are mapped to $T$ latent vectors by the encoder and the classification network then predicts the presence of disease from these $T$ latent vectors using fully connected layers.
The vectors are processed individually and are then concatenated into one vector which represents the state of the whole image sequence. 
More fully connected layers then process this vector to produce the classification.

We denote an input segmentation sequence as $\mathbf{X}= [\mathbf{x}_1, \mathbf{x}_2, ... \mathbf{x}_T] $, and its corresponding latent mean and standard deviation vectors as $\mathbf{M}=[\mathbf{\mu}_1, \mathbf{\mu}_2, ... \mathbf{\mu}_T]$
and $\mathbf{\Sigma}=[\mathbf{\sigma}_1, \mathbf{\sigma}_2, ... \mathbf{\sigma}_T]$
where $(\mathbf{\mu}_t, \mathbf{\sigma}_t) = \mathrm{Encoder}(\mathbf{x}_t)$.
The decoded images are denoted as $\mathbf{\widetilde{X}}= [\mathbf{\tilde{x}}_1, \mathbf{\tilde{x}}_2, ... \mathbf{\tilde{x}}_T]$.
During training the decoder is provided samples $\mathbf{\tilde{x}}_t = \mathrm{Decoder}(\mathbf{\mu}_t + \mathbf{\sigma}_t \odot \nu )$ where $\nu \sim \mathcal{N}(0, I)$ is a noise vector and $\odot$ denotes elementwise multiplication.
During inference, the only the mean is used and so $\mathbf{\tilde{x}}_t = \mathrm{Decoder}(\mathbf{\mu}_t)$.
The ground truth label is denoted by $y$ and the predicted label by $\tilde{y}=\mathrm{Classifier}(\mathbf{M})$.
The joint loss function for the VAE and classifier can then be written as follows:
\begin{equation}
    \mathcal{L}_{\mathrm{total}}
    = \frac{1}{T} \sum_{t=1}^{t=T} 
    \left[ \mathcal{L}_{\mathrm{recon}}(\mathbf{x}_t, \mathbf{\tilde{x}}_t) 
    + \beta \mathcal{L}_{\mathrm{KL}}(\mathbf{\mu}_t, \mathbf{\sigma}_t) \right]
    + \gamma \mathcal{L}_{\mathrm{class}}(y, \tilde{y})
\end{equation}
for constants $\beta$ and $\gamma$ which weight the components of the loss function.
$\mathcal{L}_{\mathrm{recon}}$ was chosen to be the cross-entropy between the input segmentations and the output predictions, and $\mathcal{L}_{\mathrm{class}}$ the binary cross entropy loss for the classification task.
$\mathcal{L}_{\mathrm{KL}}$ is the usual Kullback-Leibler divergence between the latent variables and a unit Gaussian, which has the effect of penalising latent vectors far from the origin and so by pulling each input's latent vector towards the origin ensures a smoothness to the latent space.
We train the model in two stages, first using only the VAE loss, i.e. $\gamma=0$, and secondly training both the VAE and classifier together using $\gamma=1$.
We set $\beta=0.2$ throughout%
, chosen by manual tuning.
The data were augmented during training by randomly applying pixel-wise shifts of up to 5 pixels in the up-down, and left-right directions.

\noindent
\textbf{Concept Activation Vectors:}
CAV \cite{Kim2017a} aim to provide explanations for a classification network's decision in terms of concepts understandable to a human.
The network is trained as usual, and the CAV analysis occurs at test time.
Data which do, and do not, contain certain human-understandable concepts are passed through the classifier, and the activations $\mathbf{z}$ at a given intermediate hidden layer are recorded.
For our experiments this layer is the fully connected layer with $64$ units, labelled `CAV' in Figure \ref{fig:model}, and the concepts are clinically relevant biomarkers measured from the segmentation.
A separate linear classifier is then trained to distinguish between the activations $\mathbf{z}$ produced by these two sets of inputs.
The CAV for a concept $c$ is the normal vector, $\mathbf{v}_c$, to this linear classifier. 
This allows an observer to measure the sensitivity of the classifier to a concept understandable to humans by evaluating the dot product between $\mathbf{v}_c$ and the gradient, at the layer of the classifier in question, of the logit value of that class, $\nabla_\mathbf{z} \tilde{y}$.
Here we apply this idea to interpret cardiac disease classification in terms of commonly used clinical biomarkers.

\section{Materials and Experiments}
\label{sec:experiments}
\begin{figure}[tb]%
\noindent\makebox[\textwidth]{
    \centering
    \subfloat{{\includegraphics[width=\textwidth]{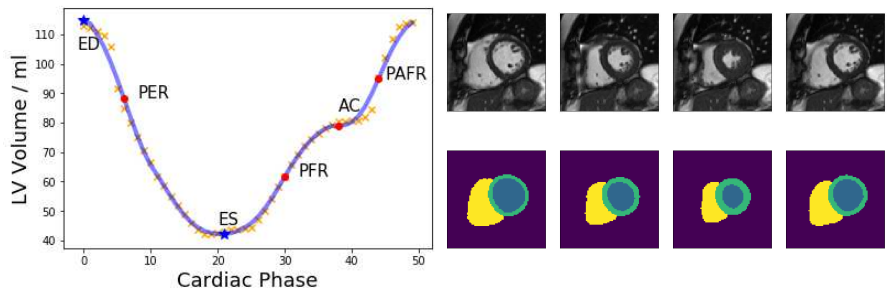}}}}%
    \caption{Left: Curve of LV volume over time, with raw data (crosses), smoothed (curve) and landmarks of the cardiac cycle annotated.
    Right: Typical cropped image sequence with 4 cardiac phases shown, and corresponding segmentations.}%
    \label{fig:example_data}%
\end{figure}
We demonstrated our approach for interpretable classification of CMR using data from the UK Biobank \cite{Petersen2016}.  
The labels for the classification task were derived from the subject's listed medical conditions according to the ICD10 disease classification.
Those listed as having any condition under I21, I22 or I25, corresponding to myocardial infarction and coronary artery disease (CAD) 
were labelled as positive.
Subjects who were labelled as negative for CAD but with other serious heart conditions (I00-I52 including hypertensive heart disease, valve disorders, congestive heart failure etc.) or who self-reported having previously had a heart attack were excluded from analysis as they could also have CAD, or very similar symptoms despite not being labelled as such.
Using the segmentation method of \cite{Bai2017} the left ventricular (LV) myocardium, blood pool, and right ventricle (RV) were each segmented as shown in the examples in Figure \ref{fig:example_data}.
%
From these segmentations we calculated several established clinical biomarkers of ventricular function which were then used as possible explanatory `concepts' in the CAV framework.
These metrics were calculated from the curve representing blood pool volume over time which was smoothed using a Savitzky-Golay filter \cite{Savitzky1964}, as shown in Figure \ref{fig:example_data}, and described in more detail in \cite{Ruijsink2019}.
The ejection fraction (EF) is defined as the fractional drop in blood pool volume from end-diastole (ED) to end-systole (ES).
The peak ejection rate (PER), peak filling rate (PFR) and peak atrial filling rate (PAFR) were determined by the magnitude of the maximal gradients of the blood pool volume over time in the relevant parts of the cardiac cycle, with the atrial contribution (AC) determined by the inflection point in this curve.
LV wall thickening was defined as the variance of LV myocardial thickening during contraction, observed between six predefined segments per image slice.
This measure indicates the presence of localised changes in myocardial contraction, and is indicative of poor cardiac health in the hypokinetic region.
\begin{table}[tb]
    \centering
\begin{tabular}{l | l |c | c}
 CAV & Description & \, $\nabla \tilde{y} \cdot \mathbf{v}_c > 0$ \, & $ \, \langle \nabla \tilde{y} \cdot \mathbf{v}_c \rangle $ \, \\
 \toprule
Low EF & Ejection Fraction & $78.2\%$ & 0.0417     \\
Low PER & Peak Ejection Rate & $88.8\%$ & 0.0770  \\
Low PFR & Peak Filling Rate & $99.6\%$ & 0.1560\\
Low APFR \, \, & Atrial Peak Filling Rate & $58.2\%$ & 0.0048 \\
High LVT & Variance of LV wall thickening \, \, & $63.4\%$ & 0.0156 \\
\end{tabular}
    \caption{The sensitivity of the classifier to clinical biomarkers of poor cardiac health.
    A biomarker with no relevance would have $\nabla_\mathbf{z} \tilde{y} \cdot \mathbf{v}_c = 0$ on average. }
    \label{tab:results}
\end{table}
A rigorous quality control process was used to remove low quality segmentations, which were typically associated with artefacts in the original images. 
Subjects with short-axis image stacks that did not cover the full LV, or intersect the apex and/or mitral valve plane were discarded.
Physiologically unrealistic segmentations were detected from the LV volume curve, determined by their having a difference of $>10\%$ in ventricular volume between the first and last segmentation in the cycle.

\begin{figure}[tb]%
  \includegraphics[width=\textwidth]{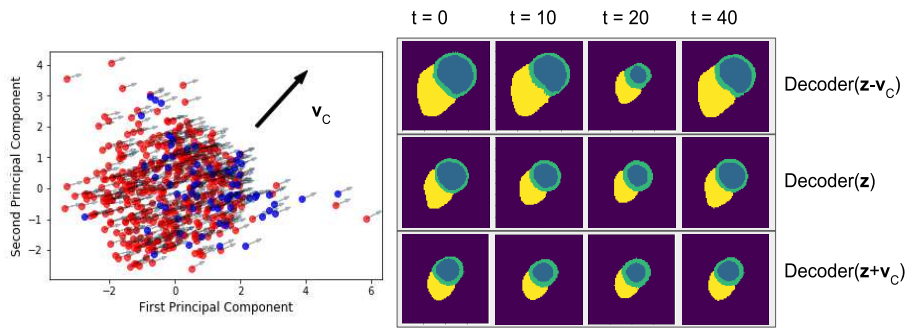}
    \caption{Left: PCA of the latent space vectors for the 500 test cases, where blue are positive for CAD.
    Each point's arrow shows the gradient of the classification logit.
    For the concept `low peak ejection rate' the CAV $\mathbf{v}_c$ is shown.
    Right: For a real test case the activations $\mathbf{z}_h$ are calculated.
    By adding $\pm \mathbf{v}_c$ and decoding the latent vectors the effect of this concept is visible in the image domain, showing noticeable changes in LV contraction.
    Four of the 50 frames in each sequence are shown here.}%
    \label{fig:latent_space_interp}%
\end{figure}

The final dataset had a total of 10,816 subjects, of which 778 were labelled with CAD.
These data were split into a training set of 5,316 subjects (708 positive, 4,608 negative), test set (70 positive, 430 negative) and a held out set of 5,000 CAD-negative subjects used for the CAV analysis.
The final trained binary classifier had an AUC of 0.78,
and the reconstructed segmentations had an average Dice Score of 0.93 with reference to the input segmentations, suggesting that the encoder and decoder networks were accurately mapping from segmentations to latent vectors and back.
The sets of images describing each human-interpretable concept were determined as follows.
In the held-out set of 5000 subjects the 1000 cases with the highest and the lowest recorded quantities of the given concept were used as the positively-labelled and negatively-labelled cases for each concept. 
To test the CAV concepts the dot product between the gradient of the classification decision with respect to the activations $\nabla_\mathbf{z} \tilde{y}$, and the concept activation vector $\mathbf{v}_c$ was measured for each case in the test set.
Table \ref{tab:results} shows the proportion of cases in which $\nabla_\mathbf{z} \tilde{y} \cdot \mathbf{v}_c > 0$, (meaning the concept had a positive impact on classification for the disease), and its mean value.
Figure \ref{fig:latent_space_interp} shows an example of a latent-space interpolation in the direction of the `low peak ejection rate' concept.
A subset of the frames are shown here but the full cardiac cycle is available in the supplementary material.
%
%
%

\section{Discussion and Conclusions}
\label{sec:discussion}
Our model not only performs classification, but also allows interpretation of features important during classification. 
Utilising CAV to interrogate the importance of well established biomarkers we found that biomarkers relating to ventricular ejection and filling rates had a large contribution suggesting that the classifier network identifies these clinically relevant features as significant. 
Latent space interpolations in the direction of the concept activation vectors, such as that in Figure \ref{fig:latent_space_interp} illustrate the ability of our method to describe its learned features, providing evidence that these vectors in the latent space correspond to typical clinical interpretations of these biomarkers.

Interpretable models do not just offer clinicians a well-calibrated estimate of the likelihood of disease.
Interpretability using known biomarkers allows the model's prediction to be placed in the context of current medical knowledge and clinical decision-making guidelines, which is a key part of translation into clinical practice.
It also has the potential to improve care by suggesting explanatory factors in an image that may have been missed or disregarded by a human.

In future work we aim to investigate which kinds of model interpretability are perceived as most informative and trustworthy by clinicians, and study the accuracy/interpretability trade-off.
We experimented with using recurrent units such as an LSTM in the classification network to process the time series of latent vectors, but found that simply concatenating them gave a superior classification performance.
Nonetheless more sophisticated architectures which more directly make use of the temporal correlations between frames should be investigated.
We also trained our model to reconstruct and predict from raw CMR images rather than segmentations.
While classification performance was comparable (AUC of 0.81) the quality of reconstructed images and latent space interpolations was not high enough (due to image blurring) to be considered usefully interpretable
.
We hope to extend our approach to the image domain using adversarial training to ensure high-quality image reconstructions which can then be used to visualise both structural and textural features relevant to the classification.
 \bibliographystyle{splncs04}

\end{document}